# Using first-principles calculations to screen for fragile magnetism: Case study of LaCrGe$_3$ and LaCrSb$_3$

Manh Cuong Nguyen[1,*], Valentin Taufour[1,2], Sergey L. Bud'ko[1,3], Paul C. Canfield[1,3], Vladimir P. Antropov[1], Cai-Zhuang Wang[1] and Kai-Ming Ho[1,3]

[1]Ames Laboratory – U.S. Department of Energy, Iowa State University, Ames, IA 50011, USA

[2]Department of Physics, University of California Davis, Davis, CA 95616, USA

[3]Department of Physics and Astronomy, Iowa State University, Ames, IA 50011, USA

*Abstract*

In this paper, we present a coupled experimental/theoretical investigation of pressure effect on the ferromagnetism of LaCrGe$_3$ and LaCrSb$_3$ compounds. The magnetic, electronic, elastic and mechanical properties of LaCrGe$_3$ and LaCrSb$_3$ at ambient condition are studied by first-principles density functional theory calculations. The pressure dependences of the magnetic properties of LaCrGe$_3$ and LaCrSb$_3$ are also investigated. The ferromagnetism in LaCrGe$_3$ is rather fragile with a ferro- to paramagnetic transition at a relatively small pressure (around 7 GPa from our calculations, and 2 GPa in experiments). The key parameter controlling the magnetic properties of LaCrGe$_3$ is found to be the proximity of the Cr DOS to the Fermi surface, a proximity that is strongly correlated to the distance between Cr atoms along the c-axis, suggesting that there would be a simple way to suppress magnetism in systems with one dimensional arrangement of magnetic atoms. By contrast, the ferromagnetism in LaCrSb$_3$ is not fragile. Our calculation results are consistent with our experimental results and demonstrate the feasibility of using first-principles calculations to aid experimental explorations in screening for materials with fragile magnetism.

## I. Introduction

Suppression of a ferromagnetic magnetic transition to zero temperature is of great interest as it may lead to quantum criticality where exotic phenomena such as non-Fermi liquid or unconventional superconductivity may emerge [1–7]. The common methods used to suppress magnetism in ferromagnetic materials includes doping or substituting magnetic element by other non- or very weak magnetic element or applying external pressure [1–5]. Doping or substitution always involve defects or randomness in the distributions of dopants and substituents making it very difficult to investigate the effects of doping/substitution on the physical properties of materials in both experimental and theoretical studies. In addition, the choices of non- or very weak magnetic dopants or substituents that do not change the crystallography of the material can be very limited or even inexistent. In contrast, external pressure is a thermodynamic parameter which is considered cleaner and can be applied to any material. Recent examples in the family of Fe-based superconductors showed that the effect of pressure on magnetism can be very similar to the effect of chemical substitutions [8–12]. This makes external pressure a very promising controlling parameter to tune physical properties of materials. For practical reasons, small or intermediate pressures are more favorable than high and ultrahigh pressure, since high and ultrahigh pressure is difficult to handle by experiment, often are not hydrostatic and hard to model, and also can induce structural transformation. A theoretical screening tool to classify whether the magnetism in a material is fragile or not under pressure will be very helpful for efficient experiment explorations [4].

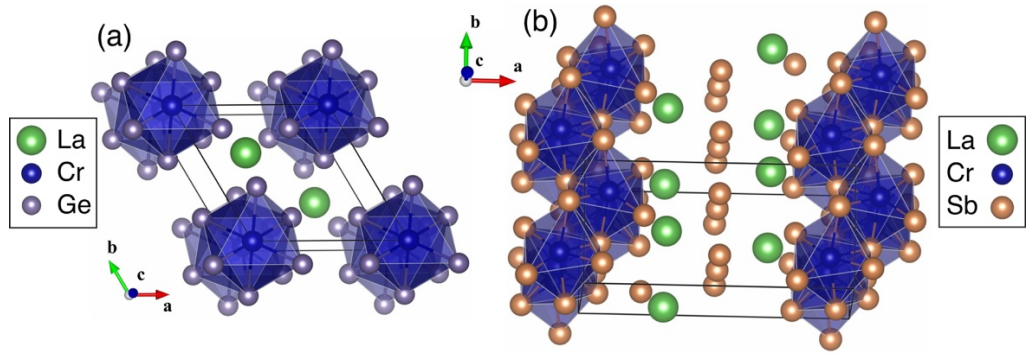

**Figure 1**. Atomic structure of (a) LaCrGe$_3$ and (b) LaCrSb$_3$. Some atoms in both structures and unit cell of LaCrSb$_3$ are repeated to show the one and two dimensional features of Cr in LaCrGe$_3$ and LaCrSb$_3$, respectively.

In this work, we demonstrate that first-principles calculations can be used as a tool to identify compounds with fragile magnetism by comparing two compounds LaCrGe$_3$ and LaCrSb$_3$ with different crystal structures (Fig. 1). LaCrGe$_3$ has a hexagonal crystal structure with P6$_3$/mmc symmetry (space group #194) and Z = 2 formula units (f.u.) in the unit cell. LaCrSb$_3$ has an orthorhombic structure with Pbcm symmetry (space group #57) and Z = 4 f.u. in the unit cell. Cr in both structures are coordinated by 6 Ge or Sb. Whereas the 6 coordinating Ge of Cr in LaCrGe$_3$ form a regular octahedron, the octahedron formed by 6 coordinating Sb of Cr in LaCrSb$_3$ is heavily distorted. The Cr-Sb bond is almost the same for 6 coordinating Sb, with less than 0.01 Å differences, but the Sb-Cr-Sb bonding angles are spreading out largely from 80 to 130° with bond angle variance $\sigma^2$ = 90.6. Another difference between these two systems is in the dimensionality of the arrangement of Cr atoms. In LaCrGe$_3$, CrGe$_6$-octahedrons form a one-dimensional line along lattice vector **c**. The CrSb$_6$-octahedrons in LaCrSb$_3$ form planes perpendicular to lattice vector **a**.

At ambient conditions, LaCrGe$_3$ shows a ferromagnetic (FM) ground state structure with magnetization aligned along the lattice vector **c**. LaCrSb$_3$ was observed to have a non-collinear magnetic configuration at low temperature [13]. There are a high moment FM coupling along the lattice vector **b** and a small moment anti-ferromagnetic (AFM) coupling along the lattice vector **c**. The magnetic moment of FM sub-lattice is 1.65 μ$_B$/f.u. and that of AFM sub-lattice is 0.49 μ$_B$/f.u. [13]. There have been several theoretical works on LaCrGe$_3$ and LaCrSb$_3$ based on density functional theory [14–18], where all of them focused on electronic and magnetic properties of materials at ambient pressure.

Our first principle calculations show that at ambient pressure LaCrGe$_3$ is a simple FM and LaCrSb$_3$ exhibits a non-collinear magnetic configuration, consistent with the experimental reports. Both systems are mechanically stable and quite compressible (especially LaCrSb$_3$) in comparison with typical metals. In the pressure dependence investigation, we first take only FM and non-magnetic (NM) phases into consideration identify the magnetism in LaCrGe$_3$ as fragile in comparison with that of LaCrSb$_3$. The key controlling parameter to the magnetic properties of LaCrGe$_3$ under pressure is the distance between Cr atoms so the effects of hydrostatic and uniaxial pressures on magnetic moment are very similar.

## II. Computational and Experimental Methods

The first-principles spin-polarized density functional theory (DFT) [19] calculations are performed using plane-wave basis Vienna *Ab-Initio* Simulation Package (VASP) [20] with projector-augmented wave (PAW) pseudopotential method [21,22] within the local density approximation parameter by Perdew and Zunger [23,24]. The energy cutoff is 450 eV and the Monkhorst-Pack's scheme [25] is used for Brillouin zone sampling. A dense k-point grid of $2\pi \times 0.025$ Å$^{-1}$ is used in all calculations, except for density of states calculations where a 1.5 times denser k-point mesh is used for accurate density of states. All structures are fully relaxed until the forces acting on each atom are smaller than 0.01 eV/Å and pressure are smaller than 0.1 GPa. The energy convergence criterion is $10^{-5}$ eV. The total enthalpy is calculated as H = E + PV, where E is the total energy, P is the external pressure and V is the system volume. The non-collinear spin-polarization and spin-orbit coupling (SOC) interaction [26] are included in some calculations for LaCrSb$_3$ at ambient pressure where noted. All other calculations are collinear spin polarized without spin-orbit coupling except where it is clearly indicated. In our experiment, single crystals of LaCrGe$_3$ and LaCrSb$_3$ were grown from solutions as reported in [2,3,5,27]. The magnetization measurements under pressure were performed using a moissanite anvil cell [28]. Daphne 7474 was used as a pressure medium [29], and the pressure was determined at 77 K by the ruby fluorescence technique [30].

## III. Results and Discussions

### 1. Physical Properties of LaCrGe$_3$ and LaCrSb$_3$ at Ambient Condition

We first discuss the structural and magnetic properties of LaCrGe$_3$ and LaCrSb$_3$ at ambient conditions. In our calculations, FM, NM and different AFM structures are studied in order to verify the experimentally observed ground structures. Specifically, for LaCrGe$_3$, we consider AFM structure along c-axis. For LaCrSb$_3$ both AFM along b-axis (b-AFM) and c-axis (c-AFM) structures are considered. We list the lattice parameters of LaCrGe$_3$ and LaCrSb$_3$ in all magnetic structures as well as from experiment and previous calculations in Table I. Our calculations confirm that the FM structure is the most stable one for both LaCrGe$_3$ and LaCrSb$_3$. In LaCrGe$_3$ system, the FM structure is -34.9 meV/f.u. lower in energy than the NM structure, which is consistent with previous calculation [18]. The AFM structure cannot be stabilized and it converged to NM structure in our calculation. For LaCrSb$_3$ within collinear magnetism, the FM structure is -48.3, -46.4 and -101.2 meV/f.u. lower in energy than the b-AFM, c-AFM and NM structures respectively, consistent with experiment and previous calculation [14]. We note that a previous calculation by Choi *et al.* [15] predicted that the FM phase is a metastable structure. Their calculations showed c-AFM is more stable than both b-AFM and FM structures.

As one can see from Table I, the LDA calculated lattice parameters of FM LaCrGe$_3$ are slightly smaller than those from experiment at 1.7 K [31], 1.4 and 2.8% smaller for a and c parameters respectively. The results are similar for LaCrSb$_3$ where LDA slightly underestimates the FM lattice parameters, within 2.2% difference from experiment at 5 K [32]. We note that DFT calculations are at 0 K but the thermal expansion effect on the lattice parameter comparison should be negligible as experimental data are collected at very low temperature as mentioned above. All previous theoretical work on LaCrSb$_3$ used experimental lattice parameters [15–17] so there is no other DFT relaxed lattice parameters to compare with our results. LaCrGe$_3$ has been synthesized in experiment recently so, apart from our work [5] there is only one theoretical work on LaCrGe$_3$ [18], which also used experimental lattice parameters.

**Table I**. Lattice parameters from experiment and calculations (T = 0 K) for different magnetic configurations of LaCrGe$_3$ and LaCrSb$_3$.

|  | LaCrGe$_3$ (Z = 2) | | LaCrSb$_3$ (Z = 4) | | |
| --- | --- | --- | --- | --- | --- |
|  | a (Å) | c (Å) | a (Å) | b (Å) | c (Å) |
| Expt | 6.165 at T = 1.7 K | 5.748 | 13.264 at T = 5.0 K | 6.182 | 6.094 |
| FM | 6.078 | 5.587 | 13.043 | 6.115 | 5.958 |
| b-AFM | N/A | | 13.003 | 6.109 | 5.948 |
| c-AFM | | | 13.003 | 6.119 | 5.955 |

**Table II**. Magnetic moment of Cr atom ($\mu_B$) and total magnetic moment of LaCrGe$_3$ ($\mu_B$/f.u.) and LaCrSb$_3$ from current (within LDA and LDA + SOC) and previous calculations. The methods used in previous calculations are also shown.

|  |  | LaCrGe$_3$ | | LaCrSb$_3$ | |
| --- | --- | --- | --- | --- | --- |
|  |  | M$_{Cr}$ | M$_{tot}$ | M$_{Cr}$ | M$_{tot}$ |
| Expt |  |  | 1.22 ~ 1.25 |  | 1.61 ~ 1.72 |
| This work | LDA | 1.17 | 1.09 | 1.65 | 1.55 |
|  | LDA+SOC | 1.20 | 1.12 | 1.70 | 1.55 |
| Previous works |  |  | 1.30 (TB – LMTO)[a] | 2.81[b,c] (LMTO) | 2.39[c] (LMTO) 2.10[d] (FPLO) |

[a]Reference [18]. [b]Reference [14]. [c]Reference [15]. [d]Reference [16].

Table II show the magnetic moment of Cr atom in LaCrGe$_3$ and LaCrSb$_3$ from experiment [2,3,13,31,33], our calculations and previously available calculations. Previous calculations were performed by linear muffin-tin orbital (LMTO) band method [15], tight-binding LMTO (TB-LMTO) [18] or full-potential local-orbital (FPLO) method [16]. In our calculations, the magnetic moment of Cr atom is 1.17 μ$_B$ in LaCrGe$_3$ and 1.65 μ$_B$ in LaCrSb$_3$. The magnetic moments of La and Ge/Sb are AFM coupled with those of Cr, so the total magnetic moment is slightly smaller than magnetic moment of Cr atom. From Table II we can see that magnetic moments are reproduced well in our LDA calculations for both LaCrGe$_3$ and LaCrSb$_3$. The total magnetic moments of LaCrGe$_3$ and LaCrSb$_3$ are slightly underestimated in comparison with experiment. Note that all previous calculations substantially overestimated the magnetic moment of LaCrSb$_3$, especially for LMTO calculation. The large overestimation of Cr moment in LaCrSb$_3$ in LMTO calculations might be due to the choice of basis set and muffin-tin radii as it is shown that different LMTO calculation settings [14,15] can give different relative stability between magnetic phases.

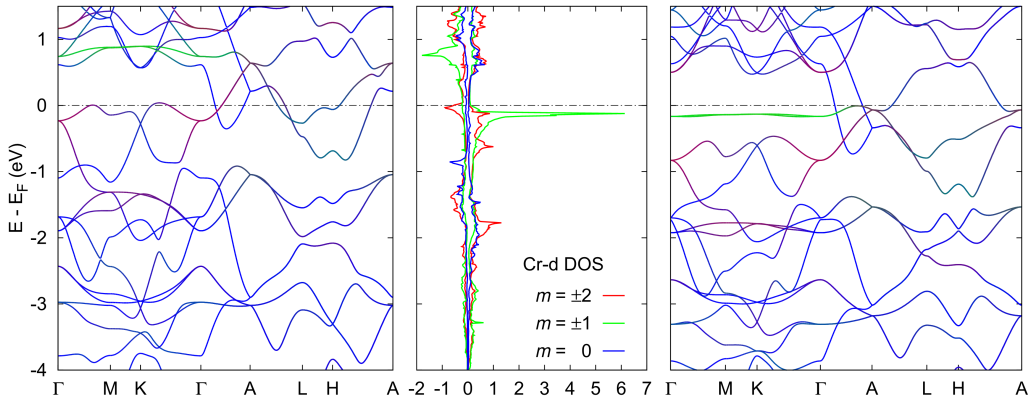

**Figure 2**. Band structures of (left) minority and (right) majority spin and (middle) density of states contributed from Cr d-orbitals in *m*-resolved of LaCrGe$_3$. Red, green and blue colors show the weight of Cr *d*-orbitals with *m* = ±2, *m* = ±1 and all other La, Cr and Ge orbitals on each band.

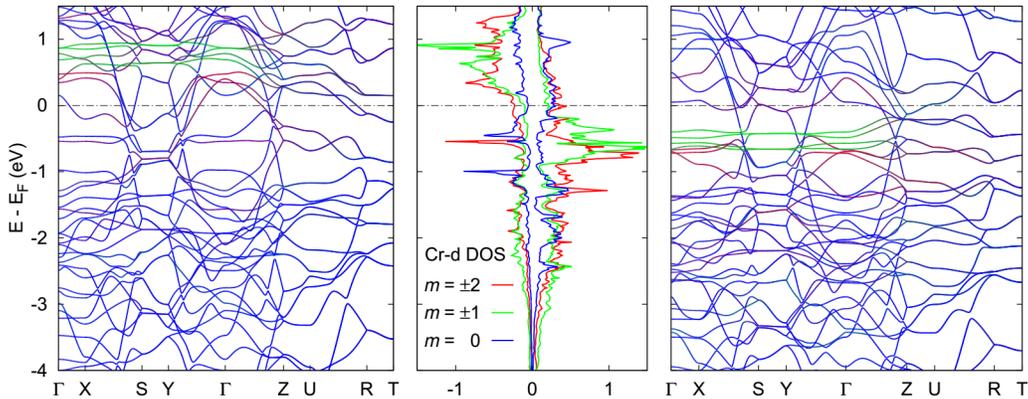

**Figure 3**. Band structures of (left) minority and (right) majority spin and (middle) density of states contributed from Cr d-orbitals in *m*-resolved of LaCrSb$_3$. Red, green and blue colors show the weight of Cr d-orbitals with $m = \pm 2$, $m = \pm 1$ and all other La, Cr and Sb orbitals on each band.

Figure 2 and 3 show band structures for majority and minority spin and projected density of states (PDOS) on Cr *d*-orbitals of FM LaCrGe$_3$ and FM LaCrSb$_3$, respectively. The positive and negative DOS values are corresponding majority and minority spin PDOS. We find that the main contributions to the density of states (DOS) near Fermi level are from Cr *d*-orbital in both systems. There are very strong Cr *d*-orbitals with m = ± 1 PDOS peaks just below the Fermi level in the majority spin channel of LaCrGe$_3$. The flat bands running through Γ-M-K-Γ segment of majority spin band structure are responsible for this strong peak. These bands are almost purely Cr d-orbital with m = ± 1. As will be seen later, the first occupied peak, count from Fermi level, of LaCrGe$_3$ is much closer to Fermi level than that of LaCrSb$_3$ and its proximity to the Fermi level is responsible for the magnetic fragility of LaCrGe$_3$. The unoccupied minority spin bands of these m = ± 1 orbitals at ~0.8 eV are also quite flat and pure as well and they are corresponding to a PDOS peak in minority spin channel at ~0.8 eV above Fermi level. There is also a small DOS peak of Cr m = ± 2 *d*-orbitals in minority spin channel very close to Fermi level.

For LaCrSb$_3$, the contribution from Cr m = ±1 and ±2 *d*-orbitals is dominating for the 1$^{st}$ main peak from -0.4 to -0.8 eV of majority spin DOS. There are very flat bands running through Γ-X-S-Y-Γ-Z segment of majority spin band structure and they are quite pure in Cr d-orbital with $m = \pm 1$. These bands induce the peaks in majority spin DOS around -0.4 and -0.6 eV. There are also quite flat band and pure in Cr d-orbital with $m = \pm 2$ going through Γ-X-S and Γ-Z segments of majority spin band structure and they are responsible for peaks near -0.8 eV. The unoccupied minority spin bands of these orbitals at ~0.9 and ~0.6 eV ($m = \pm 1$) and ~0.3 eV for ($m = \pm 2$) are also quite flat and pure as well and they are corresponding to DOS peaks in minority spin at ~0.9, ~ 0.6 and ~ 0.3 eV above Fermi level.

As mentioned above, neutron scattering experiment observed a non-collinear magnetic structure for LaCrSb$_3$ [13]. In order to investigate the effects of non-collinearity, we perform non-collinear magnetic calculation for LaCrSb$_3$ with the above relaxed lattice parameters by LDA. Since La is a heavy element, the SOC interaction is included in the non-collinear calculation. The obtained Cr magnetic moment of FM sub-lattice along b-axis is 1.66 $\mu_B$/Cr and that of AFM sub-lattice along c-axis is 0.38 $\mu_B$/Cr, which are in very good agreement with experiment. These results again show that LDA as implemented in VASP can describe the LaCrSb$_3$ system well. We also perform the non-collinear including SOC interaction calculation for LaCrGe$_3$ to investigate the effect of SOC interaction on the magnetic moment of Cr. We find that magnetic moments of Cr are collinear and the magnetic moment of Cr is enhanced slightly from 1.17 $\mu_B$ in calculation without SOC interaction to 1.20 $\mu_B$.

There have been no investigations on elastic and mechanical properties of LaCrGe$_3$ and LaCrSb$_3$. All previous work are on electronic and magnetic properties [14–18]. In this work, we also investigate elastic and mechanical properties of LaCrGe$_3$ and LaCrSb$_3$ in the thermodynamic ground states of FM. The elastic constants are calculated based on stress-strain relationship approach [34]. The Voigt (subscript V) and Reuss (subscript R) bounds of bulk (B) and shear (G) moduli are calculated by following formulas [35]:

$$B_V = (1/9)[2(C_{11}+C_{12}) + 4C_{13} + C_{33}],$$
$$G_V = (1/30)(M + 12C_{44} + 12C_{66}),$$
$$B_R = C^2/M,$$
$$G_R = (5/2)(C^2 C_{44} C_{66})/[3B_V C_{44} C_{66} + C^2(C_{44} + C_{66})],$$

where $M = C_{11} + C_{12} + 2C_{33} - 4C_{13}$ and $C^2 = [(C_{11} + C_{12})C_{33} - 2C_{13}^2)$ for hexagonal LaCrGe$_3$.

$$B_V = (1/9)[C_{11} + C_{22} + C_{33} + 2(C_{12} + C_{13} + C_{23})],$$
$$G_V = (1/15)[C_{11} + C_{22} + C_{33} + 3(C_{44} + C_{55} + C_{66} - (C_{12} + C_{13} + C_{23})]$$
$$B_R = \Delta/[C_{11}(C_{22} + C_{33} - 2C_{23}) + C_{22}(C_{33} - 2C_{13}) - 2C_{33}C_{12} + C_{12}(2C_{23} - C_{12}) + C_{13}(2C_{12} - C_{13}) + C_{23}(2C_{13} - C_{23})],$$
$$G_R = 15/\{4[C_{11}(C_{22} + C_{33} + C_{23}) + C_{22}(C_{33} + C_{13}) + C_{33}C_{12} - C_{12}(C_{23} + C_{12}) - C_{13}(C_{12} + C_{13}) - C_{23}(C_{13} + C_{23})]/\Delta + 3[1/C_{44} + 1/C_{55} + 1/C_{66}]\},$$

where $\Delta = C_{13}(C_{12}C_{23} - C_{13}C_{22}) + C_{23}(C_{12}C_{13} - C_{23}C_{11}) + C_{33}(C_{11}C_{22} - C_{12}^2)$ for orthorhombic LaCrSb$_3$.

The arithmetic averaged values of bulk or shear moduli from Voigt and Reuss bounds are the corresponding Voigt-Reuss-Hill approximation of bulk or shear moduli, respectively. Within Voigt-Reuss-Hill approximation, the Young's modulus (E) is calculated as:

$$E_H = 9B_H G_H/(3B_H + G_H).$$

The results from our calculations for elastic constants and moduli of LaCrGe$_3$ and LaCrSb$_3$ are shown in Table III. All moduli of LaCrSb$_3$ are quite smaller than those of LaCrGe$_3$. Comparing with other typical metals, moduli of LaCrSb$_3$ are comparable to those of Zn. The shear and Young's moduli of LaCrGe$_3$ are comparable to those of Co and Ni but its bulk modulus is only about two third of those of Co and Ni. From elastic constants in Table III we can verify easily that both LaCrGr$_3$ and LaCrSb$_3$ are mechanically stables as their elastic constants obey the corresponding Born mechanical stability criteria [35,36]:

$$C_{44} > 0, C_{11} > |C_{12}| \text{ and } (C_{11} + 2C_{12})C_{33} > 2C_{13}^2 \text{ for hexagonal LaCrGe}_3 \text{ or}$$
$$C_{ii} > 0 \ (i = 1,6), [C_{11} + C_{22} + C_{33} + 2(C_{12} + C_{13} + C_{23})] > 0, (C_{11} + C_{22} - 2C_{12}) > 0, (C_{11} + C_{32} - 2C_{13}) > 0 \text{ and } (C_{22} + C_{33} - 2C_{23}) > 0 \text{ for orthorhombic LaCrSb}_3.$$

**Table III.** Elastic constants of LaCrGe$_3$ and LaCrSb$_3$.

|  | LaCrGe$_3$ (GPa) | LaCrSb$_3$ (GPa) |
|---|---|---|
| $C_{11}$ | 243.8 | 133.6 |
| $C_{12}$ | 64.9 | 41.5 |
| $C_{13}$ | 58.7 | 41.7 |
| $C_{22}$ | 243.8 | 163.9 |
| $C_{23}$ | 58.7 | 34.6 |
| $C_{33}$ | 204.9 | 170.0 |
| $C_{44}$ | 80.3 | 56.0 |
| $C_{55}$ | 80.3 | 55.6 |
| $C_{66}$ | 89.5 | 47.6 |
| $B_V / B_R$ | 117.5 / 116.5 | 78.1 / 77.6 |
| $G_V / G_R$ | 84.0 / 83.7 | 55.2 / 54.4 |
| $E_H$ | 203.1 | 113.1 |

## 2. Pressure Dependence of Magnetic Moments of LaCrGe$_3$ and LaCrSb$_3$

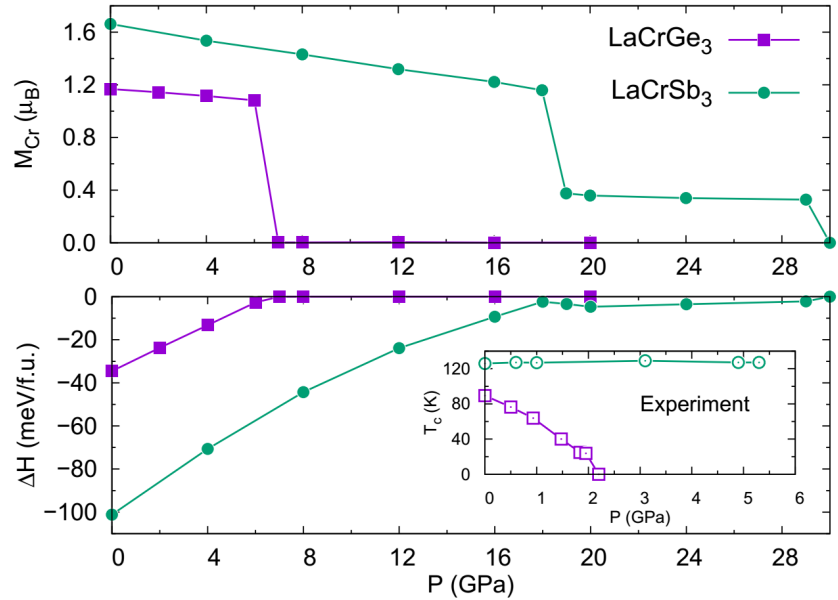

**Figure 4**. Formation enthalpy differences between FM and NM phases and magnetic moment of Cr atom as functions of pressure for LaCrGe$_3$ and LaCrSb$_3$. The inset shows the experimental results for the dependence of LaCrGe$_3$ and LaCrSb$_3$ T$_c$ on pressure [3,5].

We have also performed calculations to investigate the stability of FM phases of LaCrGe$_3$ and LaCrSb$_3$ under external pressure. The formation enthalpy differences between FM and NM phases of LaCrGe$_3$ and LaCrSb$_3$ are shown as functions of external pressure in Fig. 4. We also show in this figure the pressure dependence of magnetic moments of Cr atom in LaCrGe$_3$ and LaCrSb$_3$. The FM phases of LaCrGe$_3$ and LaCrSb$_3$ are seen to be fully suppressed by pressures of about 7 and 30 GPa, respectively. The magnetic moment of Cr in LaCrGe$_3$ is decreasing slightly with external pressure, within 10% from 1.17 μ$_B$ at ambient condition to 1.08 μ$_B$ at 6 GPa, before the NM phase becoming stable at 7 GPa. For LaCrSb$_3$, when the external pressure is increasing, the magnetic moment of Cr is decreasing more rapidly, but the FM phase will transform to another lower moment FM phase before finally transforming to NM phase. The first transition pressure, from high to low moment FM, is 18 GPa and the second transition pressure, from low moment FM to NM, is 30 GPa. The magnetic moment of the Cr atom in high moment FM phase is decreasing with pressure and it is decreasing almost linearly. For low moment FM phase of LaCrSb$_3$, there is a very small bump in the formation enthalpy difference. It increases slightly first then decreases with pressure. The formation enthalpy difference of LaCrSb$_3$ at 19 GPa is -2 meV/f.u. and it has a dip of -4 meV/f.u. at 20 GPa (Fig. 4). It should be noted that these small enthalpy differences are within the accuracy of DFT calculations. The magnetic moment of Cr atom in LaCrSb$_3$ low moment FM phase decreases slightly with external pressure between 19 and 29 GPa.

In the insert of Fig. 4 we also show the experiment results for the dependence of Curie temperature (T$_c$) of LaCrGe$_3$ and LaCrSb$_3$ on external pressure up to 6 GPa [3,5]. The experimental results show that the FM phase of LaCrGe$_3$ becomes unstable at 2.2 GPa and there is no sign of magnetic transition in LaCrSb$_3$ up to 6.0 GPa, the high pressure limit in our experiment. Our present theoretical calculation results are consistent with experiment that LaCrGe$_3$ is a fragile magnet where its ferromagnetism can be suppressed with small external pressure, whereas LaCrSb$_3$ is not fragile magnetically and requires a very high pressure to suppress its ferromagnetism. These results indicate that first-principles DFT calculation is able to predict the magnetic fragility of FM materials and it can be used to screen potential fragile FM compounds under external pressure to guide experiment explorations. However, we can see clearly an overestimation of the transition pressure from our calculation for LaCrGe$_3$. We would like to note that phase transition pressure from DFT calculation sometime can be systematically off from the experimental value, but the trend of pressure-dependent phase transition usually can be well described. For example, the predicted structural transition pressure of Si from carbon diamond (Si-I) to β-Sn (Si-II) phase is several GPa different from the experimental value, depending on exchange-correlation functional used [37,38]. But the sequence of phase transitions, *i.e.,* Si-I to Si-II to Si-V to Si-VI to Si-VII, from DFT calculations is consistent with experiment [37]. The DFT calculation, therefore, is expected to be able to predict well the magnetic phase transitions sequence. The second possible source of the overestimation of transition pressure from our calculations is the appearance of various AFM phases, so the magnetic phase transition could be from FM phase to AFM phase instead of NM phase [5]. In the current section, only FM and NM phases are taken into account. We will mention later that for LaCrGe$_3$, the appearance of AFM phases will lower the transition pressure. From Fig. 4 we also observe that for LaCrGe$_3$ there is a correlation between magnetic moment and T$_c$ as both decreasing with pressure. However, there seems to be no clear correlation between the magnetic moment and T$_c$ of LaCrSb$_3$ under pressure. The T$_c$ of LaCrSb$_3$ is almost constant with pressure up to 6 GPa in experiment.

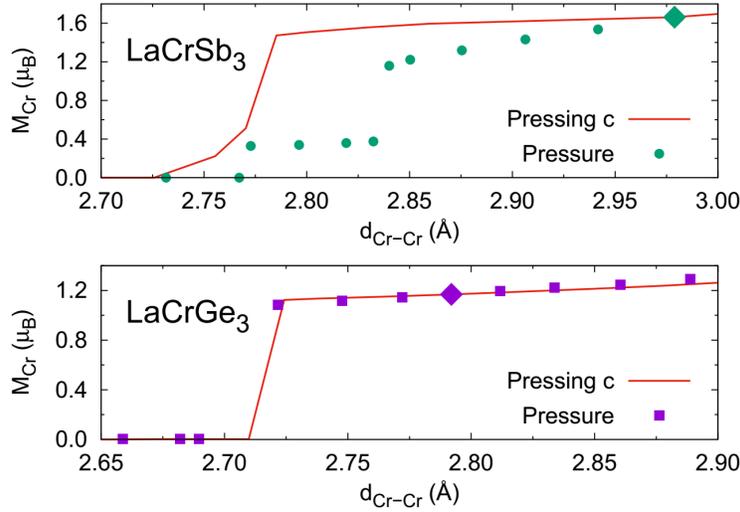

**Figure 5**. Magnetic moment of Cr as function of distance between Cr atom along c-axis of LaCrGe$_3$ and LaCrSb$_3$ under hydrostatic pressure and pressing along c-axis. Diamond marked points are for systems at ambient condition.

The Cr in LaCrGe$_3$ forms a pseudo 1-D structure along the c-axis since the distance between Cr atoms along c-axis is 2.79 Å which is much smaller than the distance between Cr atoms in the basal plane (6.078 Å). This implies that the magnetic interaction between Cr atoms are mainly between Cr atoms along the c-axis, meaning that the magnetic properties of the LaCrGe$_3$ under hydrostatic pressure and that of hypothetically uniaxially strained LaCrGe$_3$ should be very similar as long as they have the same Cr-Cr distance along the c-axis. In other words, we would see very similar ordered Cr magnetic moments for these two systems. In order to verify this, we change the length of lattice vector **c** manually and fully relax the lattice vectors **a** and **b** and internal atomic coordinates of LaCrGe$_3$ to calculate the magnetic moment of Cr. This simulation is equivalent to directional compression or uniaxial pressure experiment where the LaCrGe$_3$ system is compressed along the c-axis. Thus, experiment can be performed to verify our calculation prediction. The results of Cr magnetic moment of uniaxial pressure compressed LaCrGe$_3$ together with the results for LaCrGe$_3$ under hydrostatic pressure from above calculations as functions of the distance between Cr atoms along c-axis are shown in Fig. 5. In this figure, we also show the magnetic moment of Cr in LaCrGe$_3$ under negative pressure to compare with that of Cr in LaCrGe$_3$ under directional elongation. It is interesting that the magnetic moments of Cr in LaCrGe$_3$ under hydrostatic pressure falls almost perfectly on the magnetic moments curve of Cr in LaCrGe$_3$ under directional compression/elongation, re-confirming that Cr in LaCrGe$_3$ is indeed essentially one dimensional in terms of the magnetic interaction. On the other hand, as can be seen from Fig. 5, the magnetic moment of Cr in LaCrSb$_3$ under hydrostatic pressure and that of LaCrSb$_3$ under uniaxial pressure are resolvedly different. This is expected as we mentioned above that Cr in LaCrSb$_3$ are quasi two dimensional. For LaCrSb$_3$ at ambient condition, the distance between Cr atoms along c-axis is 2.98 Å, while the nearest distance between Cr atoms in the (a,b)-plane is 3.93 Å, which is not much larger than that along c-axis. Under the hydrostatic pressure, both the distance between Cr along c-axis and that between Cr in the (a,b)-plane of FM LaCrSb$_3$ are decreasing with pressure. Whereas the later one is increasing as we apply uniaxial pressure along c-axis on FM LaCrSb$_3$. This difference explains the difference in Cr moments shown in Fig. 5 for LaCrSb$_3$. These results show that for systems with one dimensional magnetic moment bearing atoms the effects of hydrostatic and uniaxial pressures on magnetic properties could be very close. Therefore, uniaxial pressure can be used as another method to suppress magnetism in the systems with magnetic atoms arranged in one-dimensional fashion, although experimentally the range of accessible uniaxial pressures is significantly smaller than that of hydrostatic or quasi-hydrostatic pressures. The different exotic phenomena could be emerging from magnetic materials with one dimensional magnetic atoms when hydrostatic pressure or uniaxial pressure is applied. From Fig. 5 we can roughly assign an empirical generic critical Cr-Cr distance of about 2.72 Å for the depressing of the FM phase. LaCrSb$_3$ is not a fragile magnet because the Cr-Cr distance at the ambient condition is far from this critical value, while LaCrGe$_3$ has Cr-Cr distance at ambient condition very close to the critical value so it is a fragile magnet.

Recently, further detailed investigations of resistivity and muons spin spectroscopy (μSR) [5] on the pressure phase diagram of LaCrGe$_3$ found evidence for the appearance of modulated AFM phases. The appearance of the modulated AFM phases would lower the pressure at which the FM phase being depressed since AFM phases could become more stable than the FM phase at lower pressure before NM phase become stable. Indeed, as was shown in Ref. [5], AFM orders with small q-vectors can become more stable than the FM phase and hence reduce the pressure at which the FM disappears. This demonstrate the possibility of an ideal integration between first-principles calculations and experiments: the calculations can identify which material is likely to show fragile magnetism, the experiments can discover unexpected phases which can then be considered in the calculations.

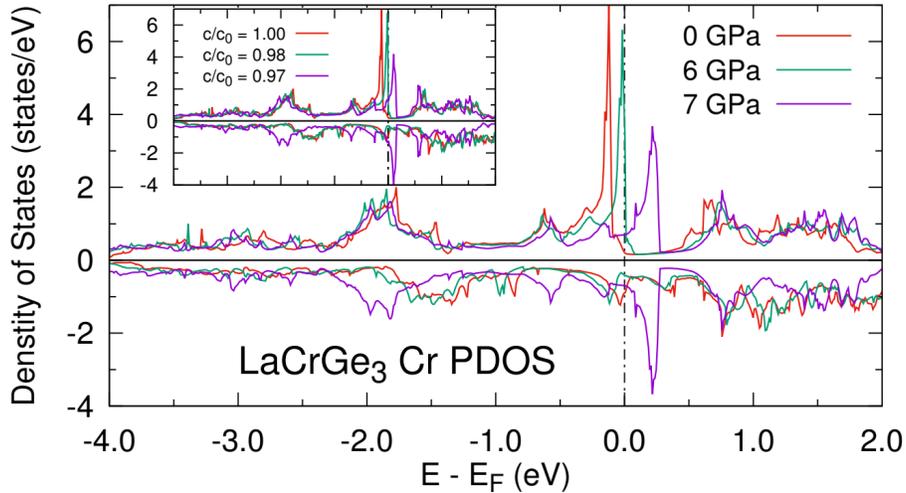

**Figure 6**. PDOS of Cr in LaCrGe$_3$ under hydrostatic pressure and (in the inset) uniaxial pressure. PDOS of NM LaCrGe$_3$ at 7 GPa and $c/c_0 = 0.97$ are plotted as half positive and half negative for better comparison.

In order to understand more physics of the transition from FM to NM phase of LaCrGe$_3$ under pressure, PDOS of Cr in LaCrGe$_3$ under different hydrostatic and uniaxial pressures are calculated. Fig. 6 shows the Cr PDOS at ambient condition and under hydrostatic pressure of 6 and 7 GPa. The figure's inset shows the Cr PDOS of LaCrGe$_3$ under uniaxial pressure along lattice vector **c**. The Cr-Cr distances in LaCrGe$_3$ structure with $c/c_0$ = 0.98 and 0.97 are very close to that of LaCrGe$_3$ under hydrostatic pressure of 6 and 7 GPa, respectively, where $c_0$ and c are the lengths of lattice vector **c** at ambient condition and under uniaxial pressure. We can see clearly from Fig. 6 that when LaCrGr$_3$ is more and more compressed under either hydrostatic or uniaxial pressure, the very high Cr PDOS peak, which is just below (about -0.15 eV) the Fermi level at ambient condition, is pushed closer and closer to the Fermi level. When this peak crosses the Fermi level, somewhere between 6 and 7 GPa or between $c/c_0$ = 0.98 and 0.97, it introduces a peculiar instability to the FM phase due to very high DOS of one spin channel at Fermi level. This instability will induce a magnetic or structural transition. We do not observe any structural transition as the symmetries are the same at 6 and 7 GPa and at $c/c_0$ = 0.98 and 0.97. The magnetic phase transition could be from FM phase to a non-collinear magnetic, an AFM or NM phase. Since we do not consider non-collinear and AFM phases in this work, what kind of magnetic phase transition cannot be resolved precisely. However, the important observation here is that the FM phase is destabilized by the high Cr PDOS peak at Fermi level. Comparing Cr PDOS of LaCrGe$_3$ at 6 and 7 GPa or between $c/c_0$ = 0.98 and 0.97 in Fig. 6, we can see a splitting of the strong Cr majority spin PDOS peak into an unoccupied peak at about 0.20 eV and other 2 occupied peaks at about -0.50 and around -1.90 eV. We obtain a similar FM to NM transition picture in LaCrSb$_3$ for the first majority spin PDOS center at ~ -0.75 eV. Thus, the proximity of a strong DOS peak to Fermi level in LaCrGe$_3$ makes the FM phase fragile and it may be used as an indicator for screening of ferromagnetic fragile materials.

**IV. Conclusions**

In summary, we have investigated elastic, mechanic, electronic and magnetic properties of LaCrGe$_3$ and LaCrSb$_3$ by first-principles density functional theory calculations. The DFT results are consistent with experiments. LaCrGe$_3$ has a ferromagnetic ground state, while LaCrSb$_3$ ground state consists of a ferromagnetic sub-lattice with moment parallel to lattice vector **b** and an anti-ferromagnetic sub-lattice with quite smaller moment parallel to lattice vector **c**. First-principles can effectively screen for fragile FM phases by taking into account FM and NM phases. When various AFM phases are considered, first-principles can predict accurately the transition pressure of FM phase of LaCrGe3 at ~2 GPa [5]. The key parameter for magnetic system with one dimensional magnetic atoms is the distance between them so the effects of hydrostatic and uniaxial pressures on magnetic moment are very similar. The proximity of LaCrGe$_3$ high DOS peak to Fermi level makes the FM phase fragile.

**Acknowledgements**

This work was supported by the U.S. Department of Energy (DOE), Office of Science, Basic Energy Sciences, Materials Science and Engineering Division, including computing time at the National Energy Research Scientific Computing Center (NERSC). The research was performed at the Ames Laboratory, which is operated for the U.S. DOE by Iowa State University under contract # DE-AC02-07CH11358. Magnetization measurements under pressure (V. T.) were supported by Ames Laboratory's laboratory-directed research and development (LDRD) funding.

*Email: mcnguyen@ameslab.gov